\documentclass[11pt]{article} 
\usepackage{hyperref} 
\pdfoutput=1 
\begin{document} 
\title{Granular Impact Dynamics: Acoustics and Fluctuations} 
\author{Abram H. Clark, R. P. Behringer \\ 
\\\vspace{6pt} Duke University, Durham, NC 27708, USA} 
\maketitle 
%% The abstract (in this file, and that submitted as text to arXiv) should include the exact phrase 
%% "fluid dynamics video" or "fluid dynamics videos" 
\begin{abstract} 
In the corresponding fluid dynamics video, created for the APS DFD 2012 Gallery of Fluid Motion, we show high-speed videos of 2D granular impact experiments, where an intruder strikes a collection of bidisperse photoelastic disks from above. We discuss the force beneath the intruder, which is strongly fluctuating in space and time. These fluctuations correspond to acoustic pulses which propagate into the medium. Analysis shows that this process, in our experiments, is dominated by collisions with grain clusters. The energy from these collisions is carried into the granular medium along networks of grains, where is it dissipated.
\end{abstract} 
Much previous work in granular impact dynamics has been \emph{macroscopic}, focusing on empirical force laws and scaling behavior. This is in part a result of difficulty of obtaining experimental data at sufficiently small space and time scales. The goal of our work is to attempt to connect macroscopic dynamics to grain-scale mechanisms. We do this with high-speed video of 2D impact experiments, where an intruder strikes a collection of approximately 25,000 bidisperse hard photoelastic disks from above. Photoelastic particles allow us to observe the local granular force response, and our typical frame rates (40,000 frames per second) allow us to resolve time scales faster than the typical time for an acoustic pulse to traverse single particle. These videos reveal rich acoustic behavior emanating from the leading edge of the intruder. We also track the intruder to determine the trajectory (depth, velocity, and acceleration), which agrees well with previous macroscopic approaches mentioned previously. By comparing the acceleration with the photoelastic activity, we show that the intruder deceleration is dominated by large force fluctuations observed in photoelastic analysis. This demonstrates a new picture for the microscale dynamics, dominated by collisions between the intruder and clusters of grains, creating acoustic pulses which carry energy away along networks of grains.

In the corresponding \href{somewhere}{video}, we show a sample photoelastic video, where a circular bronze intruder (diameter $D=20.32$~cm, bulk density which is seven times greater than the particles) strikes the granular bed from above at about 5~m/s. As the intruder moves through the granular bed, it generates substantial acoustic activity, where pulses are generated from the leading edge of the intruder, and travel away at approximately 300~m/s. We then show a side-by-side movie, where the left side is a time-averaged photoelastic movie, and the right side shows the instantaneous photoelastic response with the time-averaged frame subtracted off. The left (time-averaged) frames reveal a slowly evolving force network which is used repeatedly for many pulses. In the right (instantaneous minus average) frames, the time-averaged force network is dark, and the instantaneous granular forces manifest as bright pulses traveling along the latent grain networks.

Analysis which is not shown in the video demonstrates that it is these acoustic pulses which control the intruder deceleration. For a full analysis, please see our arXiv paper \href{http://arxiv.org/abs/1208.5724v1}{here} (submitted to PRL), or our talk at DFD 2012, titled ``What is the granular response to a high-speed impact?" (M32.00001).

% main text 
%\section{Introduction} 
%The {\em hyperref} package is used to make links to the videos. 
%%%  The format is:  \href{URL of video}{name that will appear in the text} 
%Two sample videos are 
%\href{http://ecommons.library.cornell.edu/bitstream/1813/8237/2/LIFTED_H2_EMS
%T_FUEL.mpg}{Video 
%1} and 
%\href{http://ecommons.library.cornell.edu/bitstream/1813/8237/4/LIFTED_H2_IEM
%_FUEL.mpg}{Video 
%2}. 
%It is recommended that the article include: 
%\begin{enumerate} 
%\item An explanation of what is shown in the video. 
%\item The relevant conditions, parameters, etc.. 
%\item References to any papers containing further information on the 
%videos. 
%\item In the Abstract (in the LaTeX file and in the text submitted 
%to arXiv), the exact phrase ``fluid dynamics video" or ``fluid 
%dynamics videos".  This is to facilitate subsequent searching. 
%\end{enumerate} 
%% 
\end{document}